 \documentclass[final,5p,times,twocolumn,authoryear]{elsarticle}

\usepackage{amssymb}
\usepackage{amsmath}
\usepackage{lipsum}

\journal{arXiv}

\begin{document}

\begin{frontmatter}



\title{Cooperative SIR dynamics as a model for spontaneous blood clot initiation}


\author[first]{Philip Greulich}
\affiliation[first]{organization={Mathematical Sciences, University of Southampton},
            city={Southampton},
            postcode={SO17 1BJ}, 
            country={United Kingdom}}
\affiliation[second]{organization={Institute for Life Sciences, University of Southampton},
            city={Southampton},
            postcode={SO17 1BJ}, 
            country={United Kingdom}}

\begin{abstract}
Blood clotting is an important physiological process to suppress bleeding upon injury, but when it occurs inadvertently, it can cause thrombosis, which can lead to life threatening conditions. Hence, understanding the microscopic mechanistic factors for inadvertent, spontaneous blood clotting, in absence of a vessel breach, can help in predicting and adverting such conditions. Here, we present a minimal model -- reminiscent of the SIR model -- for the initiating stage of spontaneous blood clotting, the collective activation of blood platelets. This model predicts that in the presence of very small initial activation signals, macroscopic activation of the platelet population requires a sufficient degree of heterogeneity of platelet sensitivity. To propagate the activation signal and achieve collective activation of the bulk platelet population, it requires the presence of, possibly only few,  hyper-sensitive platelets, but also a sufficient proportion of platelets with intermediate, yet higher-than-average sensitivity. A comparison with experimental results demonstrates a qualitative agreement for high platelet signalling activity.
\end{abstract}




\end{frontmatter}




\section{Introduction}
\label{introduction}
Blood clotting is the formation of macroscopic aggregates of blood platelets which are stabilised by a fibrin polymer network (coagulation) \citep{Gale2011}. The physiological purpose of such an aggregate, called a \emph{blood clot}, or \emph{thrombus}, is to seal breached vessels and to protect an individual from blood loss. However, when blood clots emerge in the blood stream, away from vessel breaches, they can occlude vessels (\emph{thrombosis}) and obstruct blood flow, which is associated with pathologies such as stroke and heart attacks \citep{Raskob2014}, leading to more than 15 million deaths per annum world-wide \citep{BHF2021}. 

The formation of a \emph{platelet plug} through platelet aggregation is the first stage of platelet-mediated blood clotting; this is then followed by the coagulation cascade that involves the development of a fibrin network to stabilise the platelet plug and form a mature blood clot \citep{Gale2011}. Platelets can only aggregate once they have been activated, either directly by agonists in the bloodstream and from vessel walls (usually arising from damaged vasculature) or via paracrine signalling from other activated platelets \citep{Stalker2012,VanderMeijden2019}. 

Physiological blood clotting occurs at vessel breaches, but blood clots can sometimes emerge without apparent vascular damage and with stimulant levels far below threshold levels expected to activate platelets (\emph{hypercoagulation}) \citep{LaPelusa2023}. Although genetic and lifestyle risk factors for this are well established, the microscopic mechanistic origin of the emergence of such \emph{spontaneous blood clotting} has so far eluded our full understanding, and quantitative modelling approaches could not reproduce or predict spontaneous blood clotting. 

It has been shown that platelets are highly diverse, and their \emph{sensitivity} -- the threshold and propensity to activate when exposed to stimulating agonists -- differs greatly between platelets within a person \citep{VanderMeijden2019,Heemskerk2022,Jongen2020}. Some platelets are extremely sensitive \citep{Baaten2017}, and it has been hypothesised that activation of such hyper-sensitive platelets -- which might be rare, but can activate at very low stimulant concentrations -- can propagate activation via paracrine signalling to activate also less sensitive platelets, thereby generating a large population of activated platelets. This could initiate platelet aggregation in situations where the abundance of stimulants is much lower than the threshold for the bulk activation of platelets \citep{Baaten2017,Jongen2020,Lesyk2019}. Thus, the diversity of platelet sensitivity, and not bulk platelet sensitivity alone, may drive platelet aggregation.

Whether activated clusters remain small or the activation signal percolates through the platelet population, leading to collective activation/aggregation, is a non-trivial critical emergent phenomenon which has not been understood yet. Although models for blood clot growth have been widely used, those established models usually model blood clot growth in the presence of a vessel breach, which secrets strong activation stimulants and where shear flow can trigger activation \citep{fogelson_phasetrans_1993,Wang1999,Fogelson2015,Fogelson2016,Schoeman2017,Link2020}. On the other hand, no model has so far been able to explain how tiny stimulant signals can lead to macroscopic activation. 

In this work, we present a \emph{minimal} mechanistic model for paracrine (platelet-platelet) activation via signalling, which captures the main qualitative features of the first phase of spontaneous blood clotting -- the macroscopic, collective activation of platelet population, upon a weak stimulant signal, which usually would not be able to activate a macroscopic platelet population alone. We thus propose a reaction-rate model which can be translated into a dynamical system for the concentration of activated and naive platelets. This system has similarities to the SIR model, a model for the spread of epidemic diseases, with the difference that multiple contagion carrying agents (here: stimulant secreting activated platelets) need to meet close to each other to spread the contagion (here: activation signal). We will show that in order to achieve macroscopic activation of the platelet population, following minor stimulation -- which from a medical point of view can lead to a dangerous thrombus that occludes otherwise undamaged vessels -- can only propagate to macroscopic proportions if there is a substantial degree of heterogeneity of platelet sensitivity.

\section{Model}

Our aim is to model the transition from a naive, non-activated population of platelets, which are not able to aggregate, to a situation where a substantial proportion of the platelet population is activated. Here, we only wish to model the first, initiating phase of thrombus formation, namely the collective activation of platelets, which allows platelets to adhere to each other and is a pre-requisite for aggregation of platelets to form a platelet plug. 

We make the following simplifying, yet biologically motivated assumptions:

\begin{itemize}
    \item The platelet population is well mixed.
    \item {\bf N}aive, non-activated platelets ($N$) can be activated either by external stimuli (e.g. collagen, convulxin), or by paracrine signalling (e.g. ADP)) from other platelets in their immediate vicinity. They require a threshold amount of signalling molecules at their surface \citep{VanderMeijden2019,Jongen2020}.
    \item Freshly {\bf a}ctivated platelets ($A_s$) {\bf s}ecret signalling molecules which can activate other platelets (``degranulation'') \citep{VanderMeijden2019}.
    \item Signalling molecules disperse quickly through diffusion and advection in sometimes turbulent blood flow \citep{Saqr2020}. Hence, the range of the signalling interaction is very short-ranged, and can be characterised by a stimulant concentration $c_0$ added by each activated secreting platelet, $A_s$, to its immediate environment.
    \item The threshold amount of stimulus required to activate a naive platelet, $c$, might be higher than $c_0$, the concentration provided by a single neighbouring secreting platelet.
    \item Degranulation occurs for a limited amount of time  \citep{Polasek2006}, after which platelets turn into an idle activated state ($A_i$), that is, they remain activated (and thus able to aggregate and initiate coagulation), but stop secreting activation signals.
    \item We consider the scenario in flow or in droplets in which shear is small. Hence, we neglect shear-flow induced activation of platelets, that plays a role for clotting at vessel walls \citep{Rana2019,Hellmuth2016}.
\end{itemize}

\begin{figure}
    \centering
    \includegraphics[width=0.9\linewidth]{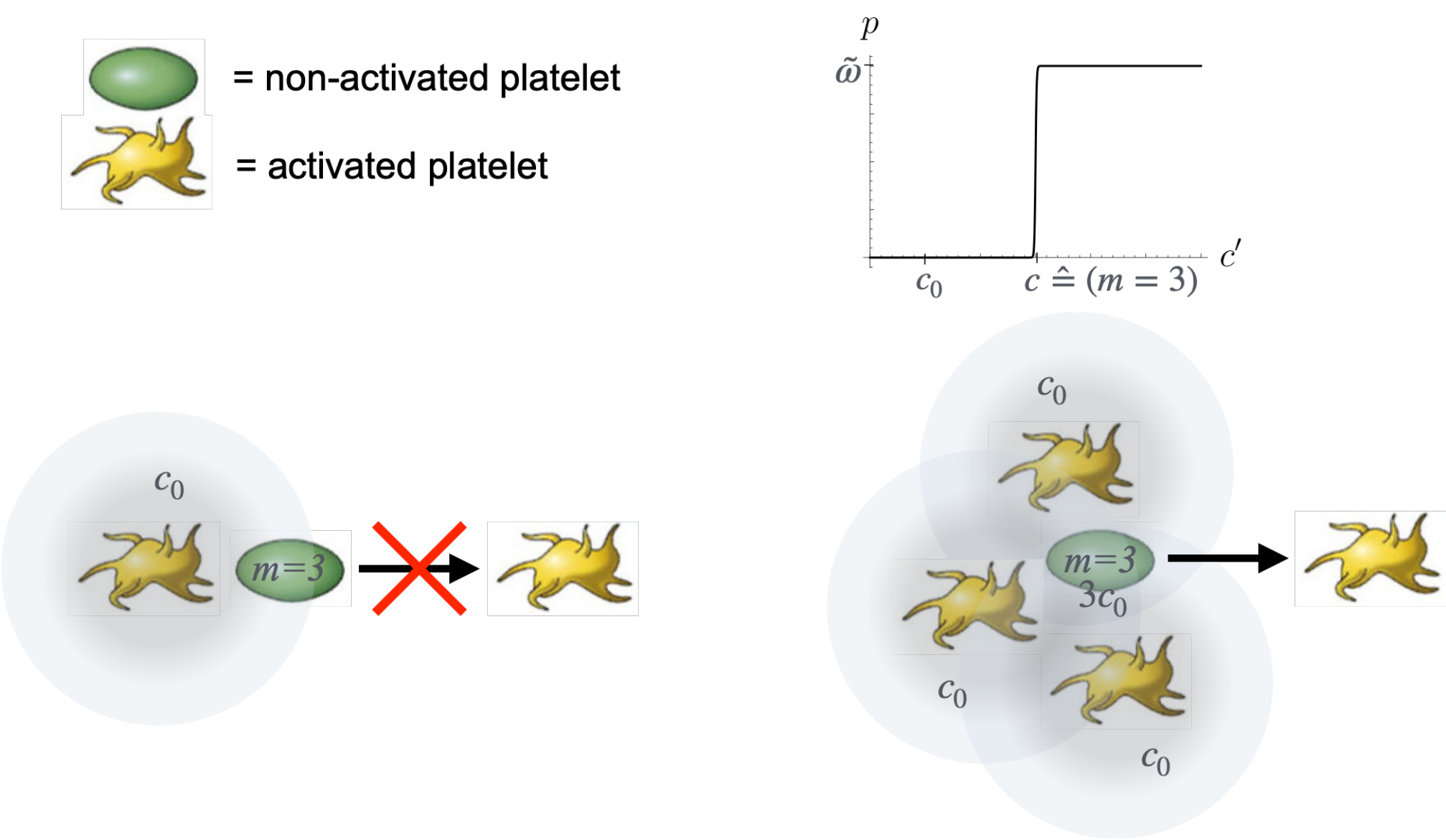}
    \caption{Illustration of the model for the case $c = 3 c_0$, meaning $m=3$. (Top right:) The propensity to activate as function of the provided stimulant concentration $c'$ for $c = 3 c_0$. Only if $c' \geq c$, platelets can activate. (Bottom:) Illustration of the cooperative activation of platelets. A single secreting activated platelet provides only a stimulant level $c_0$, not sufficient to activate other naive platelets. However, if 3 secreting platelets are close to a naive platelet, their stimulants add linearly to provide $c' = 3 c_0 = c$, which is sufficient to activate the naive platelet.}
    \label{fig:0}
\end{figure}

From these assumptions, we can formalise the activation dynamics of platelets. Let us denote the concentration of externally provided stimulant as $c_{\rm ex}$. If $m'$ secreting platelets are immediately next to a naive platelet (within the distance over which the stimulant level, $c_0$, secreted by a platelet is maintained), they generate a maximal concentration $c' = c_{\rm ex} + m' \, c_0$ in their immediate vicinity. To activate a naive platelet with threshold concentration $c$, it requires that $c' > c$, and thus,
\begin{align}
\label{m_of_c}
    m = \left\lfloor \frac{c - c_{\rm ex}}{c_0}\right\rfloor
\end{align} 
is the minimal number of secreting platelets required to activate that platelet. Hence, we can formally write the events of this model as:
\begin{align}
\label{model_events}
    N_m + m A_s \to (m+1) A_s, \quad A_s \to A_i
\end{align}
The first term describes how $m$ activated secreting platelets, $A_s$, turn a naive platelet, $N_m$, into another activated secreting platelet. The second term denotes that a secreting activated platelet stops secreting after some time and becomes an {\bf i}dle activated platelet $A_i$. These dynamics are also illustrated in Fig.~\ref{fig:0}. We note that in principle also higher numbers of meeting secreting platelets $m'>m$ can activate a platelet, but the probability for this to happen, which decreases exponentially with the number of platelets to meet, is much lower and thus neglected.

\begin{figure*}[ht]
    \centering
    \includegraphics[width=0.9\linewidth]{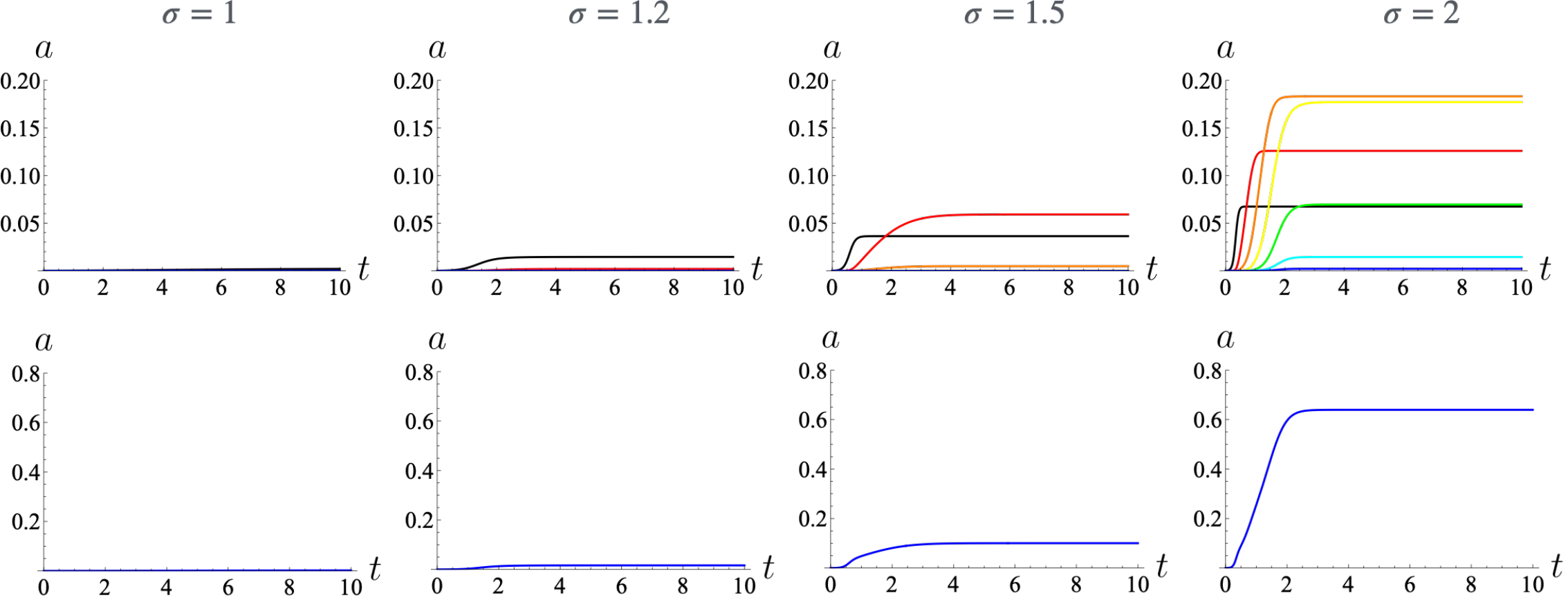}
    \caption{Time course of proportion of activated platelets, $a = a_e + a_i$ as predicted from model \eqref{coop-sir-eq_het} for $r = 300$ and platelet sensitivities being distributed normally with mean $\bar m = 4$ and different values of standard deviation $\sigma$. We have, from left to right, $\sigma = 1,1.2,1.5,2$. (Top row:) Activated platelets separated for their initial sensitivity, $a_m = a_e + a_i$, with $m=1$ (black), $m=2$ (red), $m=3$ (orange), $m=4$ (yellow), $m=5$ (green), $m=6$ (cyan), $m=7$ (blue). (Bottom row:) Total proportion of activated platelets $a = a_e + a_i$.}
    \label{fig:1}
\end{figure*}

As the population is well mixed, we can use the law of mass action to derive a dynamical model for the concentrations of naive platelets, $n_m$, secreting activated platelets, $a_s$, and idle activated platelets, $a_i$, from the ``reactions'' \eqref{model_events}:
\begin{align}
\label{coop-sir-eq}
    \dot n_m &= - \omega n_m a_s^m \\ \nonumber
    \dot a_s &= \omega n_m a_s^m - \gamma a_s \\ \nonumber 
    \dot a_i &= \gamma a_s \,\, ,
\end{align}
where $\dot x := \frac{dx(t)}{dt}$ for $x=n_m,a_s,a_i$, while $\omega$ is the specific activation rate. This rate depends on the range a secreting platelet travels during the time it keeps secreting (``degranulation time'' $t_s$ \citep{Polasek2006}), the signalling range, and the the propensity to activate $\tilde \omega$ (as in Fig.~\ref{fig:0}). The parameter $\gamma = \frac{1}{t_s}$ is the rate at which a previously activated platelet ceases secreting stimulants. 

Notably, for $m=1$, this model is equivalent to the SIR model, the paradigmatic model for the spread of a contagion \citep{KermackMckendrick1927}. In the context of infectious diseases, the case $m>1$ can be seen as a variant of the SIR model in which several infected individuals need to be present at the same time to infect others, i.e. they ``cooperate'' to spread the contagion. We therefore refer to this model, in its general form, as a \emph{cooperative SIR model}. The following analysis will study this model, and its heterogeneous version, for its general properties. Due to its relationship to the SIR model, this may have implications for understanding of infectious diseases, yet here we wish to focus on its implications for collective platelet activation and thus, eventually, for blood clotting.

To simplify the analysis, we will use a non-dimensionalised version of Eqs. \eqref{coop-sir-eq}, in which we use the degranulation time as time unit. Thus, we use rescaled time $\tilde t = \gamma t$, and the reproductive number $r = \frac{\omega\rho}{\gamma}$, where $\rho=n+a_s+a_i$ is the total concentration of platelets. Furthermore, we express the equations in terms of the proportions of subpopulations, $\tilde n = \frac{n}{\rho}, \tilde a_s = \frac{a_s}{\rho}, \tilde a_i = \frac{a_i}{\rho}$, so that $\tilde n + \tilde a_s + \tilde a_i = 1$. We can then eliminate $\tilde a_i = 1 - \tilde n_m - \tilde a_s$, and express the system by two equations. For convenience, we rename the non-dimensionalised quantities to remove the tilde: $\tilde n \to n, \tilde a_s \to n_m, \tilde a_i \to a_i, \tilde t \to t$, to arrive at the non-dimensional form of Eq. \eqref{coop-sir-eq}:
\begin{align}
    \dot n_m &= - r n_m a_s^m \label{coop-sir-eq_nd_1}\\ 
    \dot a_s &= r n_m a_s^m - a_s \label{coop-sir-eq_nd_2} \,\, .
\end{align}

We will also study a heterogeneous version of the model. First, we note that heterogeneity in $r$ is not qualitatively different to a homogeneous system. Assuming a distribution of naive platelets with different $r$, $n(r)$, we have, instead of Eq. \eqref{coop-sir-eq_nd_2}, $\dot a_s = \int_0^\infty r \, n(r) a_s^m \, dr - a_s$.  Since $a_s^m \int_0^\infty r \, n(r) dr = \bar r n a_s$, the corresponding equation is the same as \eqref{coop-sir-eq_nd_2} when replacing $r$ with the mean value $\bar r$. Thus, we do not consider this type of heterogeneity explicitly. Another type of heterogeneity is when the threshold concentration $c$ varies between platelets, according to a probability density distribution $n(c)$. Associated with this is are sub-populations $n_m$ with different threshold numbers $m = 1,2,...$, which are related to $n(c)$ by $n_m = \int_{c_{\rm ex} + m c_0}^{c_{\rm ex} + (m+1) c_0} n(c) \, dc$. Then, we get the non-dimensionalised time evolution of the system:
\begin{align}
\label{coop-sir-eq_het}
    \dot n_m &= - r n_m a_s^m \quad \mbox{ for all } m \in \mathbb N \\ \nonumber
    \dot a_s &= r \sum_m n_m a_s^m - a_s 
\end{align}
While the sensitivity of a platelet is not relevant for the dynamics after it is activated, we will for our analysis also distinguish activated platelets by sensitivity, defining the proportions of activated platelets with activation threshold $m$ as $a_m := a_s^{(m)} + a_i^{(m)} := n_m^{(0)} - n_m$ for $m=1,2,...$.

\begin{figure*}
    \centering
    \includegraphics[width=0.8\linewidth]{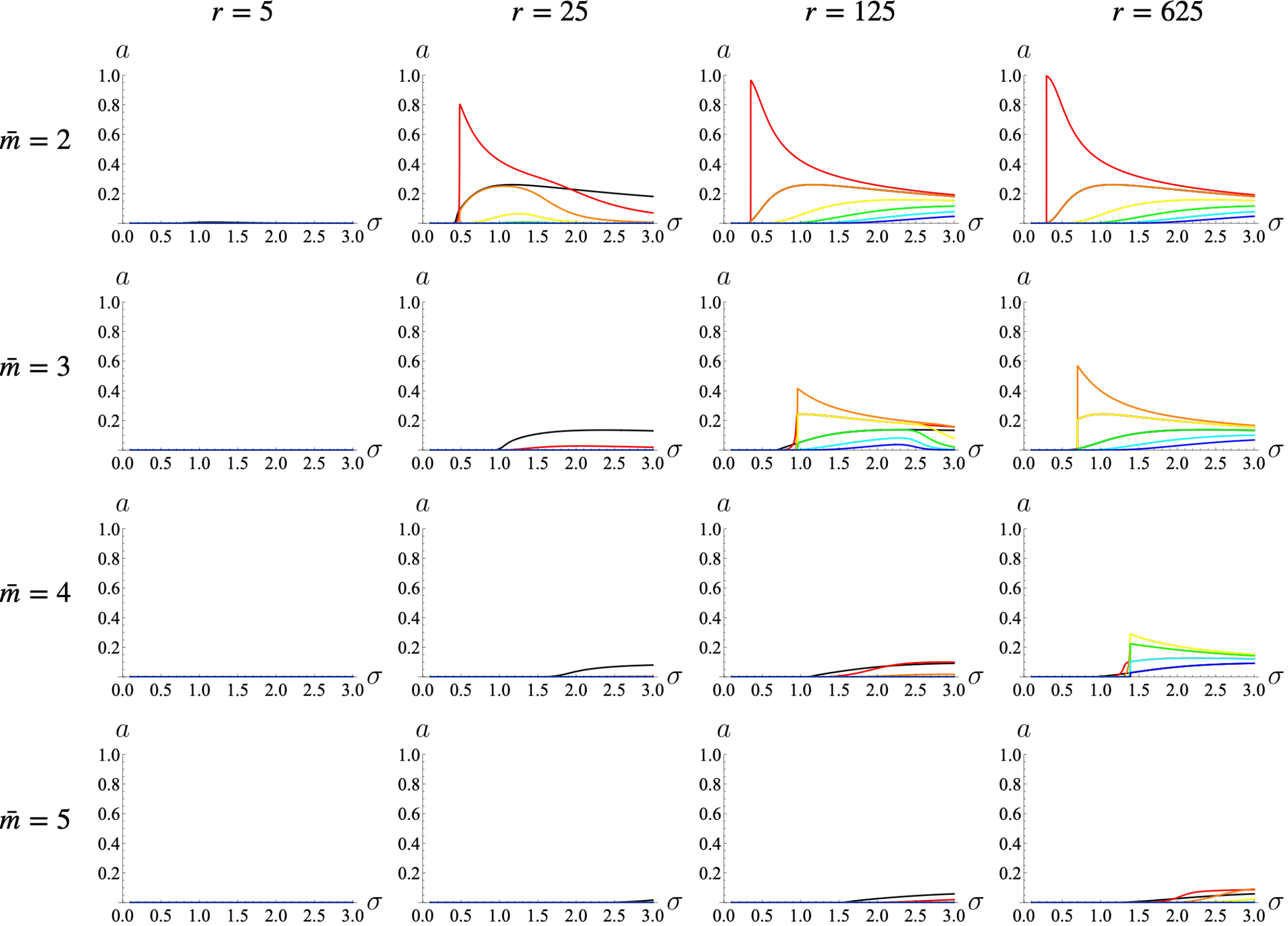}
    \caption{Equilibrium proportions of activated platelets, separated for sensitivities, $a_m = n_m^{(0)} - n_m$, as function of the width of sensitivity distribution, $\sigma$, for time $t = 10$, as predicted by model \eqref{coop-sir-eq_het}. Plots are shown for different values of the distribution's mean $\bar m$ (rows) and $r$ (columns). Rows have, from top to bottom, $\bar m = 2,3,4,5$, columns have, from left to right, $r = 5,25,125,625$. Colours represent $a_m$ for different $m$ as in Fig. \ref{fig:1}.}
    \label{fig:2}
\end{figure*}


\begin{figure*}
    \centering
    \includegraphics[width=0.8\linewidth]{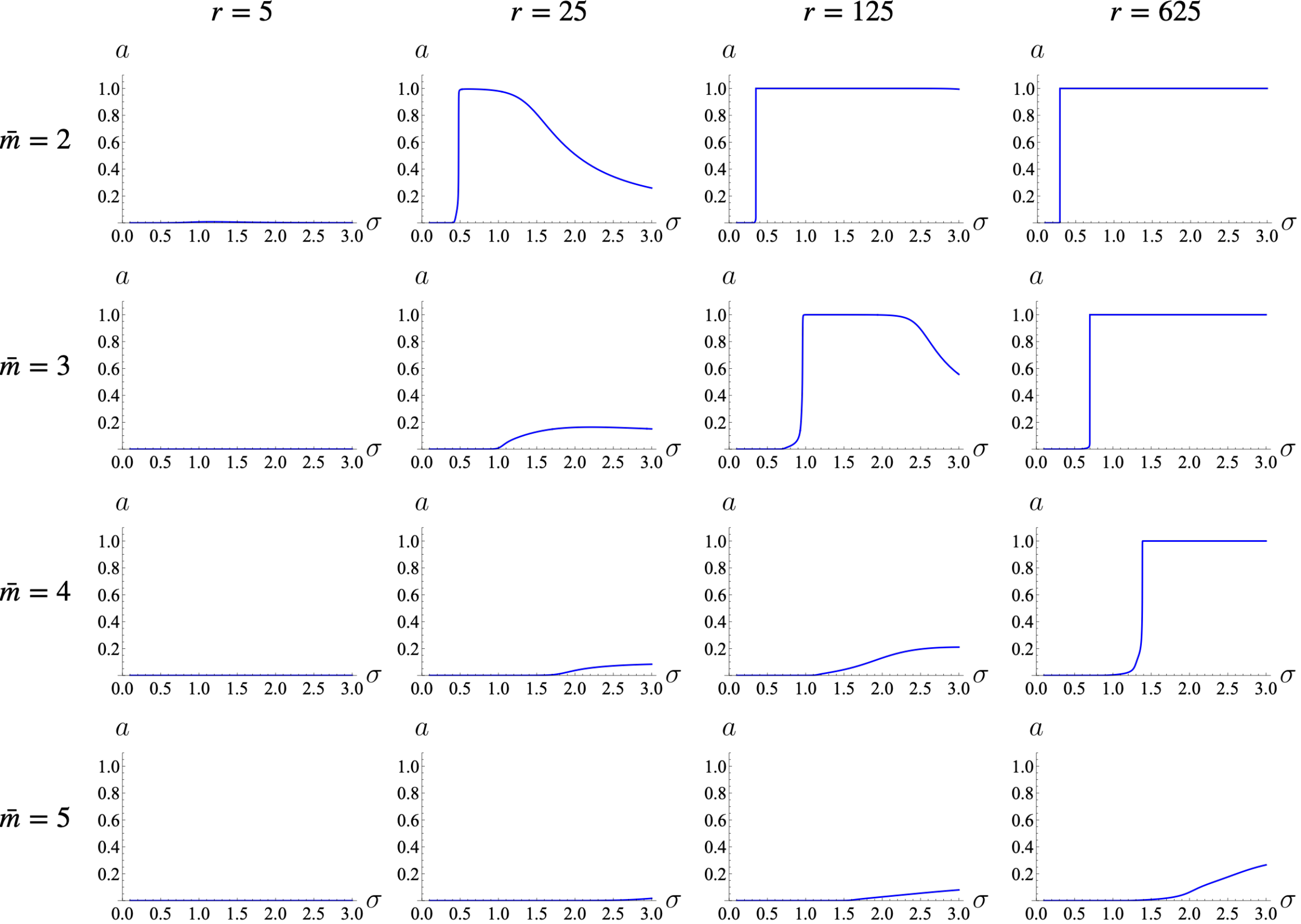}
    \caption{Equilibrium values of total proportion of activated platelets, $a = a_e + a_i$, as function of the width of (Normal) sensitivity distribution, $\sigma$, for non-dimensionalised time $t = 10$, as predicted by model \eqref{coop-sir-eq_het}. Plots are shown for different values of the distribution's mean $\bar m$ (rows) and signalling range $r$ (columns). Rows have, from top to bottom, $\bar m = 2,3,4,5$, columns have, from left to right, $r = 5,25,125,625$.}
    \label{fig:3}
\end{figure*}

\section{Results}

\subsection{Collective activation}

Our main goal is to study under which circumstances a macroscopic population of platelets activates. First, we consider the situation with diminishing amounts of external stimulant $c_{\rm ex} \approx 0$. For a convenient terminology, we define the terms ``microscopic"/``macroscopic", as follows: We consider a total of $X$ platelets (or, alternatively, stimulant molecules), which is large, that is, $X \to \infty$, and a sub-population thereof with $X_s$ platelets. The sub-population $X_s$ is \emph{microscopic}, if $X_s > 0$, but its proportion on the total population $X$, $x := \frac{X_s}{X} \to 0$ for $X \to \infty$. On the other hand, a population is \emph{macroscopic} if $X_s$ diverges for $X \to \infty$, in a way that $0<x \leq 1$. 

We assess under which circumstances a macroscopic population of activated platelets, $a := a_s + a_i = 1 - \sum_m n_m$, emerges, if initially only a microscopic population of the size $a_s = \epsilon \to 0$ of secreting activated platelets is present, if external stimulant is absent or only microscopic, $c_{\rm ex} = 0$. This is equivalent to asking under which circumstances an epidemic breaks out in a contagion model such as the SIR model. As for $m=1$, the model \eqref{coop-sir-eq_nd_1},\eqref{coop-sir-eq_nd_2} is equivalent to the SIR model, it is well known that the condition for an epidemic to break out is for $r>1$, if initially all individuals, except for a microscopic proportion, are susceptible -- which in our case corresponds to all but a few platelets being naive.

We now wish to study this question for general $m>1$ in Eqs. \eqref{coop-sir-eq_nd_1},\eqref{coop-sir-eq_nd_2} and then for the heterogeneous model, Eq. \eqref{coop-sir-eq_het}. We first note that the condition for a macroscopic platelet population to become activated upon exposure to a microscopic population of secreting activated platelets, $a_s(t=0) = \epsilon \to 0$, is equivalent to the fixed point $\boldsymbol{x}^* = (n = 1, a_s=0)$ being unstable. Thus, to assess this, we have a look at the Jacobian matrix of Eqs. \eqref{coop-sir-eq_nd_1},\eqref{coop-sir-eq_nd_2} at that fixed point for $m>1$,
\begin{align}
    J \vert_{\boldsymbol{x}^*} = \begin{pmatrix} 
    - r a_s^m & - r m n a_s^{m-1} \\
    r a_s^m & r m n a_s^{m-1} - 1
    \end{pmatrix}\vert_{(n=1,a_s=0)} = \begin{pmatrix} 
    0 & 0 \\
    0 & - 1
    \end{pmatrix} \,\, .
 \end{align}
 This matrix has eigenvalues $0$ and $-1$, therefore the fixed point is Lyapunov stable. This means that for $m>1$, no macroscopic activation of the platelet population can occur when seeded with an infinitesimally small population of activated platelets, $a_s(t=0) = \epsilon$. This has some advantages from the biological point of view: since macroscopic activation of platelets can lead to blood clotting, and inadvertent clotting can lead to dangerous pathologies through thrombosis, it could be a dangerous situation if a tiny population of such platelets were sufficient to trigger this. Hence, a situation with $m>1$ is protecting against inadvertent clotting.

 However, it has been shown that platelet sensitivity is highly heterogeneous \citep{Jongen2020}, therefore, we study under which circumstances the heterogeneous model, Eq. \eqref{coop-sir-eq_het}, with a distribution of platelet thresholds $n_m(t=0) = n_m^{(0)}$, can lead to macroscopic activation. Taking the Jacobian matrix of the heterogeneous model, Eq. \eqref{coop-sir-eq_het}, gives:
 \begin{align}
    J = \begin{pmatrix} 
    - r a_s & 0 & 0 & \cdots & - r n_1 \\
    0 & - r a_s^2 & 0 & \cdots & - 2 r n_2 a_s  \\
    \vdots & \vdots & \cdots & \vdots & \vdots\\
    r a_s & r a_s^2 & \cdots & \cdots & r \sum_m m n_m a_s^{m-1} - 1
    \end{pmatrix} \,\, ,
 \end{align}
 where the size of the matrix is $m_{\rm max} + 1$ with $m_{\rm max}$ being the largest $m$ for which $n_m^{(0)} > 0$. Taking the Jacobian at the fixed point $n_m = n_m^{(0)}$ for $m= 1,2,...$ and $a_s^{(0)}=0$, we get,
  \begin{align}
    J = \begin{pmatrix} 
    0 & 0 & 0 & \cdots & - r n_1^{(0)} \\
    0 & 0 & 0 & \cdots & - 2 r n_2 a_s  \\
    \vdots & \vdots & \cdots & \vdots \\
    0 & 0 & \cdots & \cdots & r n_1^{(0)} - 1
    \end{pmatrix} \,\, .
   \end{align}
This is a triangular matrix, which thus has the eigenvalues $0$ (with multiplicity $m_{\rm max}$) and $r \, n_1^{(0)} - 1$. As the latter eigenvalue is positive for $r \, n_1^{(0)} > 1$, we get that for those values the fixed point is unstable and thus a macroscopic proportion of the platelet population becomes activated, while for $r \, n_1^{(0)} < 1$, no macroscopic activation occurs, i.e. $a \approx 0$. For notational convenience, we thus call $n_1^* := \frac{1}{r}$ and $r^* := \frac{1}{n_1^{(0)}}$, respectively, the critical values which, when exceeded, lead to macroscopic activation. This is not surprising, since the sub-population of hyper-sensitive platelets, with $m=1$, follows the dynamics of the SIR model. In that case (when $r n_1^{(0)} > 1$), we have that $\dot a_s > r n_1 -1 > 0$ as long as $n_1 > \frac{1}{r}$. Since the activation wave ceases only once $\dot a < 0$, this means that eventually $n_1 < \frac{1}{r}$, and thus the final proportion of activated platelets, for $t=t_f \to \infty$, has a lower bound: $a = n_1^{(0)} - n_1 > n_1^{(0)} - \frac{1}{r}$. 

At first glance, it is only obvious that the sub-population of platelets with $m=1$ is activated, while it is not clear whether platelets with less sensitivity, that is, higher thresholds of $m$, will be activated to macroscopic proportions. To assess this, we consider numerical solutions of Eqs. \eqref{coop-sir-eq_het} for different distributions of platelet sensitivities and values of $r$. We first consider the scenario where platelet sensitivities are distributed according to a Normal distribution with mean activation threshold $\bar m$ and standard deviation $\sigma$. In Fig. \ref{fig:1} we see time courses of activation, separated for platelets with different $m$, with $a_m = n_m^{(0)} - n_m$ for fixed $\bar m = 4$ and different values of $\sigma$. Notably, we see that for low values of $\sigma$, no activation occurs, for larger values, activation occurs, but only for population $n_1$, while for even larger values activation of populations $n_m$ with higher $m$ occurs as well. Thus, we observe that in a heterogeneous population, the population with the highest sensitivity $m=1$, but also populations with the lower sensitivity (larger $m>1$) are activated if signalling strength $r$ and heterogeneity, characterised by $\sigma$, are sufficiently high. 

\begin{figure}[h]
    \centering
    \includegraphics[width=0.9\linewidth]{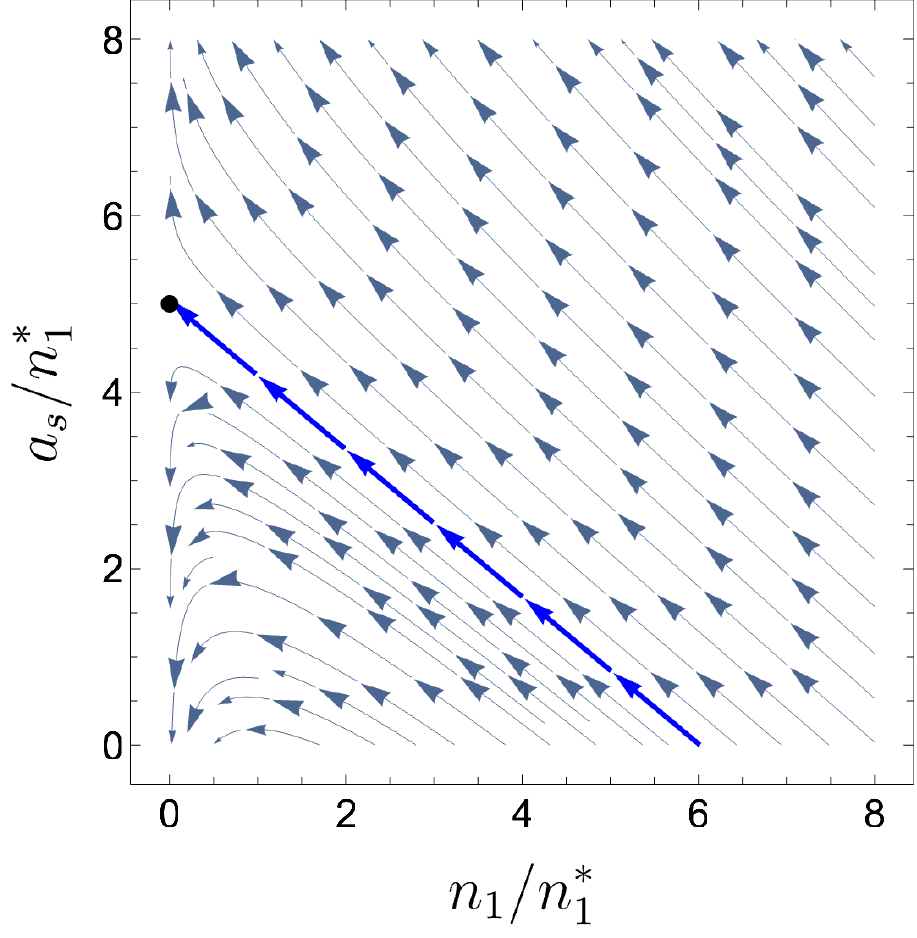}
    \caption{Phase portrait projected on $a_s,n_1$ (collection of trajectories $a_s(n_1)$), rescaled by $n_1^* = \frac{1}{r}$ for small $a_s$ when $O(a_s^3)$ can be neglected and $n_m \approx const$ for $m>1$. Parameter $r=100$. The emphasised blue trajectory shows the trajectory converging to the fixed point $(n_1 = 0,a_s = \frac{1}{n_2 r})$. This trajectory denotes the stable manifold which separates basins of attraction.}
    \label{fig:4}
\end{figure}

\begin{figure*}
    \centering
    \includegraphics[width=0.9\linewidth]{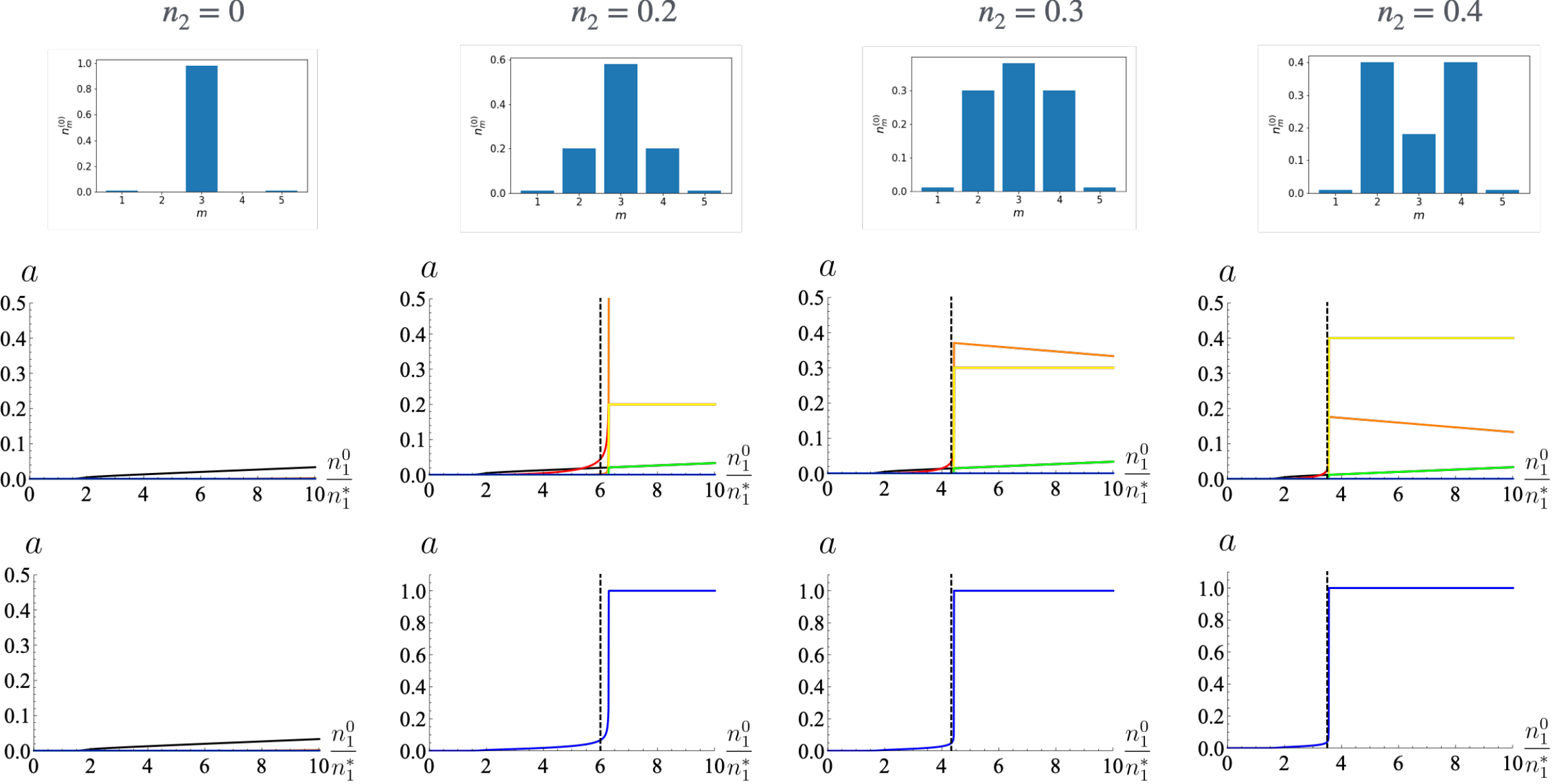}
    \caption{Equilibrium proportions of activated platelets, $a = a_s + a_i$, as function of $n_1^{(0)}$, rescaled by $n_1^* = 1/r$. The plots show different values of $n_2$ in different columns, from left to right $n_2=0,0.2,0.3,0.4$. We assume a symmetric distribution with $\bar m = 3$, which further implies $n_3^{(0)} = 1 - 2n_1^{(0)} - 2n_2^{(0)}, n_4^{(0)} = n_2^{(0)}, n_5^{(0)} = n_1^{(0)}$. The vertical dashed line indicates the critical value $n_1^{**} = n_1^{*} \frac{1+n_2}{n_2}$. (Top row:) Illustration of the distributions $n_m^{(0)}$ for $m=1,2,...,5$ used below. (Middle row:) Total activated proportion $a$. (Bottom row:) shows the activated populations separated by $m$, $a_m = n_m^{(0)} - n_m$.}
    \label{fig:5}
\end{figure*}

To further assess how the mean sensitivity and the heterogeneity of the sensitivity affect the collective activation of platelets, we show in Figs. \ref{fig:2} and \ref{fig:3} the end points of the time courses of Figure \ref{fig:1} (as the curves reach a plateau) as a function of the standard deviation of the distribution, $\sigma$. Total activated proportions $a$ (Fig. \ref{fig:3}) and individual proportions $a_m$ (Fig. \ref{fig:2}, as curves of different colours) are shown for various values of $\bar m$ and $r$. We see that for fixed values of $\bar m$, but small heterogeneity $\sigma$, there is no macroscopic activation, but as the heterogeneity increases, macrosopic activation with $a>0$ occurs. For small values of $r$, we see that when increasing $\sigma$, first only the population with $m=1$ is activated and then, for higher $\sigma$ populations with lower sensitivity ($m>1$) are activated. Notably, for higher $r$,
the activated proportions of the populations with $m>1$ exceed substantially the activated population with $m=1$. While we know that lower sensitive populations with $m>1$ cannot be activated macroscopically by microscopic $a_s$, it appears that the population with $m=1$ is initially activated, generating a non-zero proportion $a_s$, which serves as a `seed' to activate lower sensitive populations. This activation may then sustain itself if $a_s$ is large enough.

In order to understand this behaviour, we study analytically under which circumstances the population of activated platelets remains small, i.e. when $a \ll 1$ and thus $a_s \ll 1$. We wish to study this for arbitrary distributions of platelet sensitivities $n_m(t=0) = n_m^{(0)}$, $m=1,2,...$, but focus on situations where the population of hyper-sensitive platelets with $m=1$ is very small, while the other populations are much larger: $n_1^{(0)} \ll 1$ and $n_m^{(0)} \gg n_1^{(0)}$ are such that $\frac{n_1^{(0)}}{n_m^{(0)}} \sim O(a_s) \ll 1$ for all $m>1$. Furthermore, since macroscopic activation can only occur for $n_1^{(0)} > \frac{1}{r}$ we additionally assume that $n_1^{*} = \frac{1}{r} \ll 1$, that is, $r$ is large. This is also a biologically realistic regime: as the degranulation time is around 60s \citep{Polasek2006}, which is roughly the time blood circulates once through the entire circulatory system, any secreting platelet can potentially get in contact with a very large number of naive platelets and activate them.

Since, according to these assumptions, $a_m \ll 1$, we can assume $n_m \approx n_m^{(0)}$ being approximately constant for $m>1$, and it is therefore sufficient to consider only $n_1,a_s$ as dynamical variables. We note that the first two terms of Eq. \eqref{coop-sir-eq_het}, $r n_1 a_s$ and $r n_2 a_s^2$ are both of order $O(a_s^2)$ while the other terms are of order $O(a_s^3)$, thus we have, in leading order of (small) $a_s$ for Eq. \eqref{coop-sir-eq_het}:
\begin{align}
\label{coop-sir_quadr-terms}
    \dot n_1 &= - r n_1 a_s \quad \\ \nonumber
    \dot a_s &= r n_1 a_s + r n_2 a_s^2 - a_s + O(a_s^3) \approx r n_1 a_s + r n_2 a_s^2 - a_s
\end{align}
We can now study the fixed points of this system when neglecting $O(a_s^3)$.  We see that there are fixed points for any $a_s=0$, and there is one fixed point for $a_s > 0$, namely for $n_1=0$ and $a_s = \frac{1}{n_2 r} =: a_s^*$. A linear stability analysis (see Appendix A) shows that this is a saddle point, which means that there is a stable manifold, which, since this is a two-dimensional system, consists of the trajectories that converge to the point $\boldsymbol{x}^* = (n_1=0,a_s=\frac{1}{n_2 r})$. This stable manifold separates the phase space into two basins of attraction, one which converges to $a_s = 0$ and one where trajectories diverge. To illustrate this, we show the phase portrait of system \eqref{coop-sir_quadr-terms} for $r=100$ in Fig. \ref{fig:5}. Here, we see that the highlighted trajectory separates the other trajectories according to their fate. If $n_1^{(0)}$ lies below this curve, $a_s$ cannot exceed $a_s^* = \frac{1}{n_2 r}$. On the other hand, if it lies above that curve, $a_s$ will diverge, meaning that our approximation will break down and thus $a_s \gg \frac{1}{n_2 r}$. 

To find the stable manifold which separates basins of attraction, we express trajectories as functions $a_s(n_1)$ in the $n_1$-$a_s$-plane. They can be found as solutions to the differential equation,
\begin{align}
    \frac{da_s}{dn_1} = \frac{\frac{da_s}{dt}}{\frac{dn_1}{dt}} \approx -1 -\frac{n_2}{n_1} a_s(n_1) + \frac{1}{r n_1} \,\, .
\end{align}
This is a linear ODE, whose solution can be found, for example, through the integrating factor method. As we are looking for a trajectory converging to $(n_1=0,a_s=\frac{1}{n_2 r})$, we can choose the initial condition $a_s(n_1=0) = \frac{1}{n_2 r}$, and with this we get the solution,
\begin{align}
    a_s(n_1) = \frac{1+n_2 - r n_1 n_2}{r n_2(1+n_2)} \,\, .
\end{align}
Furthermore, since $\dot n_1 < 0$, the direction of this trajectory is towards $\boldsymbol{x}^*$, hence this is indeed the stable manifold. Now, we need to distinguish whether $n_1^{(0)}$ is below or above that trajectory. To this end, we determine the intersection of this trajectory with the $n_1$ axis, which we find as the solution to $0 = \frac{1+n_2 - r n_1 n_2}{r n_2(1+n_2)}$, which is $n_1^{**} = \frac{1 + n_2}{r n_2}$. Hence, for $n_1^{(0)} < n_1^{**}$, $a_s$, and thus $a$, remain below $a_s^* = \frac{1}{n_2 r}$, while for $n_1^{(0)} > n_1^{**}$, $a_s$, and thus $a$, exceed this value substantially. Note that in the case $r\to \infty$, $\frac{1}{n_2 r} = a_s^*$ must be microscopic, so the condition for $a$ remaining microscopic is $n_1^{(0)} < \frac{1 + n_2}{r n_2}$. In that case, we thus get that substantial proportions of $a_m$ emerge once $n_1^{(0)}$ exceeds $n_1^{**} = n_1^* \frac{1+n_2}{n_2}$. This is shown in Fig. \ref{fig:5} to be an accurate prediction: there final activation $a$ is shown as function of $n_1$ for different values of $n_2$ and a mean value $\bar m=3$ \footnote{Symmetry of the sensitivity distribution then implies $n_3^{(0)} = 1 - 2 (n_1^{(0)} + n_2^{(0)}), n_4^{(0)} = n_2^{(0)}, n_5^{(0)} = n_1^{(0)}$}. 

This demonstrates that a microscopic proportion of hyper-sensitive platelets, $n_m^{(0)}$, is sufficient to trigger a macroscopic activation also of platelet populations with less sensitivity, $m>1$, including those with average sensitivity $\bar m$. Nonetheless, for $n_2^{(0)} = 0$, no substantial activation is observed. Thus, we conclude that the presence of platelets with intermediate sensitivity, between hyper-sensitivity with $m=1$, and the mean sensitivity $\bar m$, are essential for macroscopic activation. Although we do not have analytical means to study further terms in Eq. \eqref{coop-sir-eq_het}, we expect that more populations with $m>2$ can further amplify the spread of the activation signal if $n^{(0)}_m \gg n^{(0)}_{1}$, as has been indicated above in Figs. \ref{fig:1}-\ref{fig:3}.

\subsection{Comparison with experiments with external stimulants}

We now consider the situation where an external stimulant is present which is able to directly activate platelets, $c_{\rm ex} > 0$. According to Eq. \eqref{m_of_c},  the threshold number of secreting platelets $m$ required to activate platelets with threshold concentration $c$ is $m = \lfloor \frac{c - c_{\rm ex}}{c_0} \rfloor$. Hence, given a probability density distribution $n(c)$ of platelets with threshold concentrations $c$, we get a distribution of initially naive platelets of $n_m = n_{\left\lfloor\frac{c + c_{\rm ex}}{c_0}\right\rfloor}$. This means that with increasing $c_{\rm ex}$ the distribution of platelets gets shifted towards lower $m$. Importantly, platelets with $c < c_{\rm ex}$ are activated from the beginning, when the stimulant is given. We thus have as initial condition $a_s(t=0) = \int_{c' < c_{\rm ex}} n(c') \, dc' \approx \sum_{m < \frac{c_{\rm ex}}{c_0}} n_{m}$. Hence, with increasing $c_{\rm ex}$, both the proportion of initially secreting activated platelets increases, and the distribution of platelets is shifted towards lower $m$, consequently enhancing activation.

We wish to compare our model results with experimental results from droplet microfluidics assays with platelets, performed by \cite{Jongen2020}. In one of those experiments, platelets were enclosed in micro-droplets and exposed to the external stimulant convulxin, at varying concentration, and then the abundance of the activation marker P-selectin was measured via fluorescent cytometry, that is, the fluorescent intensity, $I$, was recorded (Fig. 2B in \citep{Jongen2020}). Two scenarios are compared: (A) where each droplet contains a single platelet, (B) where droplets contain platelet collectives. In situation (A) no platelet can activate another platelet through paracrine signalling since they are separated by droplets, hence $c = c_{\rm ex}$ and thus the proportion of activated platelets at a given concentration of convulxin is the cumulative distribution $F(c) = a_s(t=0) = \int_{c' < c} n(c') dc'$. Hence, the distribution of sensitivities can directly be inferred from this data. In scenario (B), platelets can activate each other, and thus this measures the collective activation of a platelet population. This can be compared with the predicted proportion of activated platelets, $a$, from our model.

However, our model has limitations modelling this data, as the experimental setup does not meet all assumptions of our model. In particular, since platelet collectives are encapsulated in a flow-free confined environment (droplets), the agonists secreted by platelets are not quickly dispersed, and thus can accumulate and reach platelets which are even further away. This means that we can expect an enhanced paracrine activation between platelets, resulting in a higher value of $r$ than would be expected in an in vivo situation. Nonetheless, we wish to test whether the \emph{qualitative} behaviour, namely the predicted collective activation of the population through a small, hyper-sensitive population of platelets, prevails. To that end, rather than finding accurate precise numerical model fits, we explore the parameter space to find values for which the model qualitatively reproduces the data. 

To determine an estimate for the distribution $n(c)$ and thus $n_m$, we fit a cumulative distribution function to the data of single-platelet droplets (A). First, we need to renormalise the experimental fluorescence measurements, $I$, from \cite{Jongen2020}, as there is always some background signal even in absence of activation. Hence, we define the \emph{relative activation} as $a_{\rm exp} = \frac{I-I_{\rm min}}{I_{\rm max} - I_{\rm min}}$ where $I_{\rm min}$ and $I_{\rm max}$ are minimum and maximum fluorescence values. As the data in \citep{Jongen2020} shows a symmetric sigmoidal form on a logarithmic scale of $c$, we will attempt to fit a log-normal distribution, with mean $\bar c$ and logarithmic variance $\sigma_c^2$ to the data. For an easier comparison with our model, which uses $m = \frac{c}{c_0}$ to define platelet populations, we will try only values $\bar c = \bar m c_0$ and $\sigma_c = \sigma_m c_0$ with integer $\bar m, \sigma_m = 1,2,...$. To that end, we first need to find an estimate for $c_0$. As for $c_{\rm ex} = c_0$ the sub-population with $m=1$ is being activated for the first time upon increasing values of $c_{\rm ex}$, we identify $c_0$ as the concentration for which, upon increasing concentrations of $c_{\rm ex}$, for the first time substantial collective activation emerges in data (B). While due to the lack of intermediate data points, this estimate can only be a rough one, we can estimate this from inspection of data (B) to be around $c_0 = 0.3$ ng/mL. While some activation is present already at $c_{\rm ex} = 0.1$ ng/mL, this could be due to spontaneous activation that is expected from our model according to our previous discussion.

\begin{figure}
    \centering
    \includegraphics[width=0.7\linewidth]{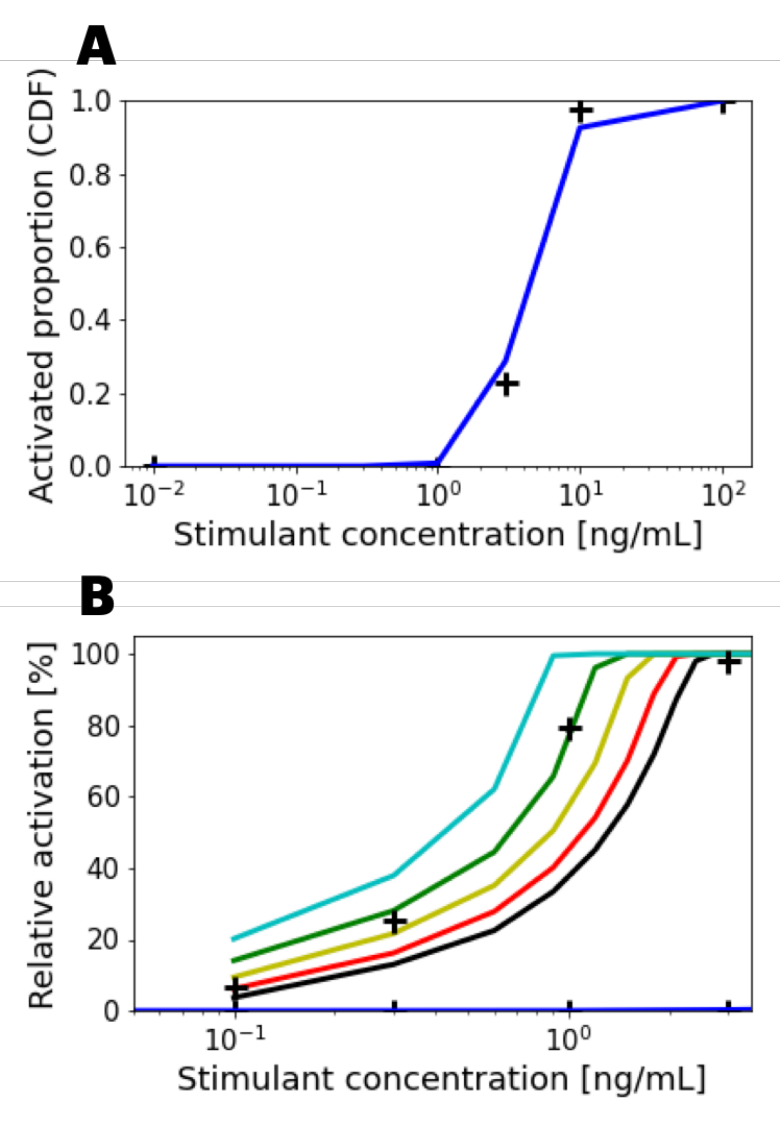}
    \caption{Comparison with experimental results from \citep{Jongen2020}. Pluses show experimental results measuring overall P-selectin activity (a marker for platelet activation) as fluorescence levels via flow cytometry, rescaled for background fluorescence and as proportion of maximal fluorescence (see text). Error bars are smaller than the symbol sizes. Curves show model predictions for $a = a_e + a_i$. The horizontal axis shows varying concentrations of stimulant (convulxin) to which platelets were exposed in the experiments, the y-axis shows relative activation. (A) Shows a log-normal distribution with mean $c=4.2$ ng/mL and standard deviation $\sigma = 0.6$ ng/mL, together with experimental results where single platelets are encapsulated in single droplets. This is a proxy for the cumulative distribution $F(c) = \int_{c'<c} n(c') \, dc'$. (B) Shows model output for $a$, compared with experimental results where many platelets are in single droplets, for different values of $r$: $r=1000$ (black), $r=3000$ (red), $r=10000$ (orange), $r=30000$ (green), $r=100000$ (cyan).}
    \label{fig:6}
\end{figure}

We find that for $\bar m = 14$ and $\sigma_m = 2$, corresponding to $\bar c = 4.2$ ng/mL and $\sigma_m = 0.6$ ng/mL, a good match of the log-normal distribution with the single-platelet droplet data, (A), is achieved, as is shown in Fig. \ref{fig:6}A. We will thus use $n^{(0)}_m = \mathcal{N} e^{-\frac{(\ln m - \ln \bar m)^2}{2 \sigma_m^2}}$ as the initial distribution of platelet activation thresholds in absence of external stimulant, where $\mathcal{N}$ is a normalisation factor that accounts for the discreteness of the values $m$. Then, we determine the initial distribution of platelet sensitivities with external stimulant $c_{\rm ex}$ as $n_m(t=0) = n^{(0)}_{m + c_{\rm ex}/c_0}$ and $a_s(t=0) = \sum_{m<\frac{c_{\rm ex}}{c_0}} n_m^{(0)}$, and we solve the system \eqref{coop-sir-eq_het} numerically using Scipy's \verb|solve_ivp| function. The result is shown in Fig. \ref{fig:6}B for different values of $r$, together with the experimental data. We note that for smaller values of $r$ no reasonable match is achieved, but for very high values of $r$, in particular for $r = 30000$ (green curve in Fig. \ref{fig:6}B), a good qualitative agreement is achieved. This is consistent with the expectation that due to the accumulation of paracrine signalling molecules within droplets, a single platelet can reach and activate a much higher population of platelets than would be in a (possibly turbulent) \emph{in vivo} flow environment. Overall, this demonstrates that our model is able to qualitatively reproduce the experimental data on collective platelet activation, though at parameter values which are likely beyond the corresponding \emph{in vivo} values.

\section{Conclusions}

Blood clotting is an important physiological process, but when it occurs inadvertently, it can cause thrombosis which can be live threatening. Thus, understanding and eventually predicting blood clotting can aid in preventing thrombosis, while retaining the physiological features of blood clotting. The initiating stage on the path towards blood clotting is the collective activation of platelets, hence understanding the platelet activation dynamics will aid in assessing blood clotting risk. 

We have introduced a minimal model for the activation dynamics of blood platelets, which includes activation by external stimuli and through stimulation via paracrine, platelet-platelet signalling. In particular, we considered the scenario in blood flow, where due to fast dispersion of stimulants, one or more activated platelets that secret activation signals are required in the immediate vicinity of a non-activated, naive platelet, to activate the latter. From these and other biologically motivated assumptions, we formulated a minimal dynamical model, \eqref{coop-sir-eq}, for the time evolution of naive and activated platelets, with the aim to predict the proportion of activated platelets when starting with a predominantly naive platelet population. If the number of secreting platelets required to activate a naive one, $m$, is equal to one, this model is equivalent to the SIR model -- the paradigmatic contagion model -- and macroscopic activation occurs whenever the 'reproductive number' $r$ is above a critical value, even for microscopic proportions of initially activated secreting platelets. For $m>1$ the model, which we then call \emph{cooperative SIR model}, still has the characteristic of a contagion model, yet, if the population is uniform, no macroscopic activation, of a non-zero proportion of a large platelet population, is possible, if initially only few platelets are secreting. 

Nonetheless, it is known that platelet sensitivity varies considerably between platelets, therefore we considered a heterogeneous version of the model where $m$ may differ between platelets. We showed that in this case, a macroscopic activation of the population is possible when only few activated platelets are initially present, if $r$ is sufficiently high. In that case, the \emph{hyper-sensitive} sub-population with $m=1$ becomes activated, and this activated, secreting sub-population then serves as a 'seed' to activate sub-populations with higher $m$, in a staggered cascade of activating ever less sensitive sub-populations (with higher $m>1$). Notably, we showed analytically and numerically that a microscopic proportion of platelets with $m=1$ is sufficient to achieve this, if $r$ is sufficiently large. This supports the hypothesis that a rare population of hyper-sensitive platelets could be able to mediate macroscopic activation and thus trigger blood clotting \citep{Baaten2017,Jongen2020,Lesyk2019}. We further showed that in order to activate the bulk of the platelet population, also platelets with less, yet higher-than-average sensitivity, with thresholds $m$ between that of hyper-sensitive and average platelets (thresholds $1 < m < \bar m$), are required to achieve macroscopic activation of the bulk platelet population. This is biologically required to then achieve macroscopic platelet aggregation and eventually blood clotting. A 'gap' in sensitivities, for example if there are no, or only few platelets with $m=2$, may break the activation wave, and no substantial activation of the platelet population as a whole emerges.

Finally, we compared our model results with experimental results where platelet activation and aggregation was measured with platelets encapsulated in micro droplets \citep{Jongen2020}. While this experimental setup did not meet all our model assumptions -- in particular since secretory, stimulating signals can accumulate -- we showed that our model is capable of qualitatively reproducing the observed dose-response curves, namely the activation intensity as function of provided external stimulant, if the reproductive number $r$ is very large. 

As most clinical platelet function tests for assessing thrombosis risk measure bulk platelet properties to assess thrombosis risk \citep{Anghel2020,Pabinger2009}, these findings suggests that clinical practice may need to be revised: if the distribution of (higher-than average) platelet sensitivities is determining the onset of blood clotting, rather than bulk platelet sensitivity, then measurements of the latter -- for example via droplet microfluidics as done by \cite{Jongen2020} -- would be required for appropriate assessment of the thrombosis risk.

To summarise, we showed that a simple model, reminiscent of the SIR model, is able to reproduce measured qualitative features of collective platelet activation. This model explains mechanistically how even very rare hyper-sensitive platelets can serve as an activation seed that may propagate the activation through the whole population to achieve a macroscopic activation, if sufficient amounts of less, but larger-than-average sensitivities are present. Once a substantial fraction of the platelet population have been activated, they may then aggregate and form the body of a blood clot. This model serves as a first step towards a development of a comprehensive model for whole blood clotting, that may allow quantitative predictions for blood clotting probability, and hence the risk of thrombosis.


\section*{Acknowledgements}
I thank Jonathan West for guidance in processing the experimental data of \citep{Jongen2020}.

\appendix

\section{}
We wish to analyse the stability and stable manifold of the fixed point of Eq. \eqref{coop-sir_quadr-terms}. This system has fixed points for any $a_s = 0$ and for $a_s = \frac{1}{n_2 r}$ and $n_1 = 0$. The Jacobian of the system at this fixed point is:
\begin{align}
    J = \begin{pmatrix} 
    - r a_s & - r n_1 \\
    r a_s & r n_1 + 2 r n_2 a_s - 1
    \end{pmatrix}\vert_{n_1=0,a_s = 1/r} =
    \begin{pmatrix} 
    - 1 & 0 \\
    1 & 1
    \end{pmatrix}
 \end{align}
This matrix is in triangular form and thus has eigenvalues $-1$ and $1$. Hence it is a saddle point, which possesses a stable trajectory, which separates the space into to basins of attraction, defined as the trajectory coverging to the fixed point. In the main text, we show that $a_s(n_1) = \frac{1+n_2 - r n_1 n_2}{r n_2(1+n_2)}$ is this trajectory.

\bibliographystyle{elsarticle-harv} 
\bibliography{library}






\end{document}